\documentclass[twocolumn,showpacs,preprintnumbers,amsmath,amssymb,superscriptaddress]{revtex4}

\newtheorem{theorem}{Theorem}
\newtheorem{definition}{Definition}

\begin{document}

\title{A Family of Concurrence Monotones and its Applications}  
\author{Gilad Gour}\email{ggour@math.ucsd.edu}
\affiliation{Theoretical Physics Institute, University of Alberta, 
Edmonton, Alberta T6G 2J1, Canada}
\affiliation{Department of Mathematics, University of California/San Diego, 
        La Jolla, California 92093-0112}

\date{\today}

\begin{abstract} 
We extend the definition of concurrence into a family of entanglement
monotones, which we call concurrence monotones. 
We discuss their properties and advantages as computational manageable measures of 
entanglement, and show that for pure bipartite states all measures of entanglement 
can be written as functions of the concurrence monotones.
We then show that the concurrence monotones 
provide bounds on quantum information tasks. As an example,
we discuss their applications to remote entanglement distributions (RED)
such as entanglement swapping and remote preparation of bipartite 
entangled states (RPBES). 
We prove a powerful theorem which states what kind of (possibly mixed) 
bipartite states or distributions of bipartite states can not be
remotely prepared. The theorem establishes an upper bound on the amount of
$G$-concurrence (one member in the concurrence family)
that can be created between two single-qudit nodes of quantum networks
by means of tripartite RED. For pure bipartite states the 
bound on the $G$-concurrence can always be saturated by RPBES.
\end{abstract}

\pacs{03.67.-a, 03.67.Hk, 03.65.Ud}

\maketitle

\section{Introduction}

Entanglement is one of the main ingredients of non-intuitive
quantum phenomena. Besides of being of interest from a fundamental point of 
view, entanglement has been identified as a non-local resource for quantum 
information processing~\cite{NC}. In particular, shared bipartite entanglement is a crucial 
resource for many quantum information tasks
such as teleportation~\cite{Ben93}, quantum cryptography~\cite{BB84}, 
entanglement swapping~\cite{Zuk93}, 
and remote state preparation (RSP)~\cite{Ben01,Shi02,Leu03,Ye04} that are employed
in quantum information protocols. 

One of the remarkable discoveries on bipartite entanglement, is that for pure states, 
there is a unique and single measure of entanglement, called 
entropy of entanglement~\cite{BBPS}, that quantifies, {\em asymptotically}, 
the non-local resources of a large number of copies of a pure bipartite state. 
However, the generalizations of the entropy of entanglement to mixed states yields,
even asymptotically, more than one measure of entanglement, 
such as entanglement of formation and distillation~\cite{BVSW}.
Despite the enormous efforts that have been made
in the last years, mixed entanglement lacks a complete 
quantification~\cite{Horodecki}. 

For a finite number of shared pure states, the entropy of entanglement is not
sufficient, and more measures of entanglement are
required to quantify completely the non-local resources. These are called 
{\it entanglement monotones}~\cite{Vidal} since they behave
monotonically under local transformations of the system. 
The family of entanglement monotones $E_{k}$ ($k=0,1,2,...,d-1$)
which introduced in~\cite{Vidal2} 
were first defined over the set of pure states as
\begin{equation}
E_{k}(|\psi\rangle)=\sum_{i=k}^{d-1}\lambda _{i}\;,
\label{one}
\end{equation}
where $\lambda_{0}\geq\lambda_{1}\geq\cdots\geq\lambda_{d-1}$ are the 
Schmidt numbers of the $d\times d$-dimensional bipartite
state $|\psi\rangle$, and then extended to mixed states by means of the convex roof 
extension. For a pure state $|\psi\rangle$ these measures of entanglement quantify 
{\em completely} the non-local resource since all the Schmidt coefficients of 
$|\psi\rangle$ are determined by them. The entanglement monotones defined
in Eq.~(\ref{one}) play a central role in
transformations of pure states by local operations and classical 
communications (LOCC)~\cite{Vidal2,Nielsen,Jon99}. Moreover, each member of
the family may quantify the possibility to perform a particular task in 
quantum information processing (for example, $E_{2}=1-\lambda _{0}$ quantifies the 
possibility to perform faithful teleportation with partially entangled states~\cite{Gou04}).  

Nevertheless, the family of entanglement monotones $E_{k}(\rho)$ is not enough to 
quantify completely the entanglement of a bipartite mixed state $\rho$. Furthermore, 
it will be argued here, that if $\rho$ is a $d\times d$-dimensional mixed state with $d>4$, in
general, it is impossible to find analytical expression (i.e. an explicit formula like 
in~\cite{HW97,Woo98}) 
for $E_{k}(\rho)$ (as well as for the entanglement of formation and other measures 
of entanglement). 
Thus, we are motivated to look for other sets of monotones which are 
more computationally manageable.

Such a computationally manageable measure of entanglement is the {\em concurrence}.
The concurrence as a measure of entanglement
was first introduced in~\cite{HW97,Woo98} for an entangled pair of qubits and later on 
generalized to higher dimensions~\cite{Run01,Min04} (there are other generalizations
of concurrence which we will not discuss here~\cite{Uhl00}).
Already in~\cite{HW97,Woo98} the importance of the concurrence monotone was recognized
and the entanglement of formation of a mixed entangled pair of qubits was calculated 
explicitly in terms of the concurrence. In higher dimensions there is not yet
an explicit formula for the generalized concurrence~\cite{Run01}, 
but lower bounds have been found~\cite{Min04}.
Recently, it has been shown~\cite{GS04} that the concurrence
plays also a major role in remote entanglement distributions (RED) protocols such as
entanglement swapping (ES) and remote preparation of bipartite entangled
states (RPBES).   

In this paper we introduce a family of entanglement monotones which
we call {\it concurrence monotones}. We discuss its properties and show 
that for pure states {\em all} measures of entanglement can be written
as functions of the concurrence monotones. We show that these concurrence monotones 
can serve as a powerful tool to rule out the possibility of certain tasks in quantum 
information processing. In particular, we find an upper bound on the entanglement 
that can be produced by tripartite RED protocols and show that the
protocol given in~\cite{GS04} for RPBES saturates the bound. 
The measure of entanglement is taken to be 
one of the members in the concurrence family, which we give the name $G$-concurrence, since
for pure states the $G$-concurrence is the {\em Geometric mean} of the Schmidt numbers.
In addition, we provide an operational interpretation of the $G$-concurrence as a type
of entanglement capacity.
 
This paper is organized as follows. In section II we define the family of concurrence 
monotones and then discuss its importance and advantages. In section III we discuss 
its applications to RED protocols and in section IV we summarize our results and 
conclusions.    

\section{Definition of concurrence monotones}

In the following, we will use the definition of concurrence as given 
in~\cite{HW97,Woo98} for the $2\times 2$ dimensional case, and its generalization 
to higher dimensions as given in~\cite{Run01} (see also~\cite{Min04}). 
The concurrence of a pure bipartite normalized state $|\psi\rangle$ 
is defined as
\begin{equation}
C\left(|\psi\rangle\right)\equiv\sqrt{\frac{d}{d-1}\left(1-{\rm Tr}\hat{\rho}_{r}^{2}\right)}\;,
\label{con}
\end{equation}
where the reduced density matrix $\hat{\rho}_{r}$ is obtained by tracing over
one subsystem. In the definition above we added the factor $\sqrt{d/(d-1)}$ so that 
$0\leq C\left(|\psi\rangle\right)\leq 1$. For $d=2$ Eq.~(\ref{con}) also coincides with 
the definition given in~\cite{HW97,Woo98} by means of the ``spin flip'' 
transformation. The concurrence of a mixed state, $\hat{\rho}$, is then defined as 
the average concurrence of the pure states of the decomposition, minimized over
all decompositions of $\hat{\rho}$ (the convex roof): 
\begin{equation}
C\left(\hat{\rho}\right)=\;{\rm min}\;\sum_{i}p_{i}C\left(|\psi_{i}\rangle\right)
\;\;\left(\hat{\rho}=\sum_{i}p_{i}|\psi _{i}\rangle\langle\psi _{i}|\right)\;.
\label{conm}
\end{equation}
In the following definition of the family of concurrence monotones, the concurrence
defined in Eqs.~(\ref{con},\ref{conm}) is denoted by $C_{2}$ since it is the
second member of the family.

\begin{definition}
(a) Consider a $d\times d$-dimensional bipartite pure state $|\psi\rangle $ with Schmidt 
numbers $\lambda\equiv(\lambda_{0},\lambda _{1},...,\lambda_{d-1})$.
The $d$ {\em concurrence monotones}, $C_{k}(|\psi\rangle)$ ($k=1,2,...,d$), 
of the state $|\psi\rangle $ are defined as follows~\footnote{See also~\cite{Bar01,Fan03} for slightly different 
definitions}:
\begin{equation}
C_{k}(|\psi\rangle)\equiv\left(\frac{S_{k}\left(\lambda_{0},\lambda _{1}...,\lambda_{d-1}\right)}
{S_{k}\left(1/d,1/d,...,1/d\right)}\right)^{1/k}\;,
\label{CM}
\end{equation}
where $S_{k}(\lambda)$ is the $k$th elementary symmetric function of $\lambda_{0},\lambda _{1},...,\lambda_{d-1}$.
That is,
\begin{align}
& S_{1}(\lambda)=\sum_{i}\lambda _{i}\;,\;\;S_{2}(\lambda)=\sum_{i<j}\lambda _{i}\lambda _{j}
\;,\nonumber\\
& S_{3}(\lambda)=\sum_{i<j<k}\lambda _{i}\lambda _{j}\lambda _{k}\;,\;
 ...,S_{d}(\lambda)=\prod _{i=0}^{d-1}\lambda _{i}.
\label{sym}
\end{align}
(b) Consider a $d\times d$-dimensional bipartite mixed state $\rho $. The $d$ {\em concurrence monotones}, $C_{k}(\rho)$, 
of the state $\rho $ are then defined as the average $C_{k}$ of the pure states of the decomposition, minimized over all 
decompositions of $\rho $ (the convex roof):
\begin{equation}
C_{k}(\rho)={\rm min}\sum_{i}p_{i}C_{k}(|\psi _{i}\rangle)\;\;\left(\rho=\sum_{i}p_{i}|\psi _{i}\rangle\langle\psi _{i}|\right)\;.
\label{CMM}
\end{equation}
\end{definition}

The functions $S_k(\lambda)$ and $[S_k(\lambda)]^{1/k}$ are Schur-concave 
(see p.78,79 in~\cite{Majo}). Moreover, 
\begin{equation}
S_k(\lambda)\leq S_k(1/d,1/d,...,1/d)=\frac{1}{d^{k}}{d\choose k}\;, 
\end{equation}
since the vector $(1/d,1/d,...,1/d)$ is majorized by {\em all} vectors 
$\lambda=(\lambda _{0},...,\lambda_{d-1})$ with non-negative components 
that sum to 1. Thus, $0\leq C_{k}(|\psi\rangle )\leq 1$ and 
$C_{k}(|\psi\rangle )=1$ only when all the Schmidt numbers of 
$|\psi\rangle $ equal to $1/d$ (i.e. $|\psi\rangle $ is a maximally
entangled state).

Eq.~(\ref{CM}) together with the convex roof extension of $C_k$ to mixed states
(see Eq.~(\ref{CMM})) defines an entanglement monotone for each $k$. To see that,  
first note that
\begin{equation} 
C_k(|\psi\rangle )=
f_k\left({\rm Tr}_{{}_{B}}|\psi\rangle \langle\psi|\right)\;,
\end{equation}
where the trace is taken over one subsystem (say Bob's system) and 
$f_k(\sigma)\equiv [S_k(\lambda(\sigma))/S_k(1/d,...,1/d)]^{1/k}$ ($\lambda(\sigma)$
is the vector of eigenvalues of the density matrix $\sigma$). According to Theorem 2 in~\cite{Vidal}
$C_{k}$ is an entanglement monotone if $f_k(\sigma)$ is a unitarily invariant,
concave function of $\sigma$. The concavity of $f_k(\sigma)$ follows from   
two facts. First (see p.79 in~\cite{Majo}),
for any two vectors $x$ and $y$ with $x _{i},y _{i}\geq 0$ 
($i=0,1,...,d-1$)
\begin{equation}
\left[S_k(x+y)\right]^{1/k}\geq [S_k(x)]^{1/k}+[S_k(y)]^{1/k}\;.
\end{equation}
Second, for two Hermitian matrices $A$ and $B$, 
$\lambda(A+B)\prec \lambda(A)+\lambda (B)$ (see p.245 in~\cite{Majo}). 
Thus, given two density matrices $\sigma _{1}$ and $\sigma _{2}$
we have ($0\leq t\leq 1$)
\begin{align}
&f_k[t\sigma _1+(1-t)\sigma _2]  =
\left[{S_k\left[\lambda\Big(t\sigma _1+(1-t)\sigma _2\Big)\right]\over 
S_k(1/d,...,1/d)}\right]^{1/k}\nonumber\\
& \geq \left[{S_k\left[\lambda\big(t\sigma _1\big)+\lambda\big((1-t)\sigma _2\big)\right]
\over 
S_k(1/d,...,1/d)}\right]^{1/k}\nonumber\\
& \geq \left[{S_k\left[\lambda\left(t\sigma _1\right)\right]\over 
S_k(1/d,...,1/d)}\right]^{1/k}+\left[{S_k\left[\lambda\big((1-t)\sigma _2\big)\right]\over 
S_k(1/d,...,1/d)}\right]^{1/k}\nonumber\\
& = tf_k(\sigma _1)+(1-t)f_k(\sigma_{2})\;.
\end{align}
Thus, Eqs.~(\ref{CM},\ref{CMM}) define entanglement monotones.

\subsection*{Advantages of concurrence monotones}

There are several advantages and applications for these particular measures
of entanglement. First, the family of concurrence monotones as defined in 
Eq.~(\ref{CM},\ref{CMM}) is {\em complete} in the sense that all the Schmidt coefficients 
of a given pure state can be determined by the $d$ concurrence monotones.
To see that, let us define the characteristic polynomial 
$f_{\lambda}(x)=(x-\lambda_{0})(x-\lambda _{2})\cdots (x-\lambda _{d-1})$
whose singular values are the Schmidt numbers. It is easy to see that 
$f_{\lambda}(x)$ can be written as
\begin{equation} 
f_{\lambda}(x)=
\sum_{k=0}^{d}\frac{(-1)^{k}}{d^{k}}{d \choose k}x^{d-k}\left[C_{k}(\lambda)\right]^{k}\;,
\label{char}
\end{equation}
where $C_{k=0}(\lambda)\equiv 1$ and $C_{k=1}(\lambda)\equiv \sum_{i}\lambda _{i}=1$.
Hence, the singular values of $f_{\lambda}(x)$ (i.e. the Schmidt numbers) are determined 
completely by the concurrence monotones $C_{k}$.

Furthermore, consider a pure $d\times d$-dimensional state
\begin{equation} 
|\psi\rangle=\sum_{ij}a_{ij}|i\rangle_{A}|j\rangle _{B}\;, 
\label{state}
\end{equation}
where
$|i\rangle_{A}$ and $|j\rangle_{B}$ are some $d$-dimensional bases in 
Alice and Bob systems, respectively. The Schmidt numbers are the non-zero
eigenvalues of the matrix $A^{\dag}A$ (or $AA^{\dag}$), where the matrix elements of
$A$ are $a_{ij}$. Thus, in general, for $d>4$, according to Abel's impossibility 
theorem (also Galois) there is no analytical expression for the Schmidt numbers
in terms of $a_{ij}$. The advantage of our family of concurrence 
monotones is that one can always express analytically $C_{k}(|\psi\rangle)$
in terms of $a_{ij}$:
\begin{equation}
C_{k}(|\psi\rangle)=d\left[\frac{{\rm Tr}B^{(k)}}{{d\choose k}}\right]^{1/k}\;,
\label{123}
\end{equation}
where $B^{(k)}$ is the $k$th compound of the matrix $A^{\dag}A$ (see p.502 
in~\cite{Majo} for the definition of compound matrices). Such an explicit formula
(in terms of $a_{ij}$) is not available for most of the measures of entanglement
discussed in literature (including the entropy of entanglement, $\alpha$-entropy or
Renyi entropy, and the family of entanglement monotones given in~\cite{Vidal2}). 

As an example, consider the entropy of entanglement 
$E(|\psi\rangle)=-{\rm Tr\rho_{r}\log\rho_{r}}$, where 
$\rho_{r}\equiv {\rm Tr} _{{}_{B}}|\psi\rangle \langle \psi|$
is the reduced density matrix. If $|\psi\rangle$ is given in terms
of $a_{ij}$ as above, then in order to calculate the entropy of entanglement,
one must be able to write $\rho_{r}$ in its diagonal form. However,
for $d>4$, in general, it is impossible to solve the equation $f_{\lambda}(x)=0$
analytically ($f_{\lambda}(x)$ is defined in Eq.~(\ref{char})).  

For $d\leq 4$ the entropy of entanglement can be expressed
in terms of the concurrence monotones. 
For $d=2$, the solution to the quadratic equation $f_{\lambda}(x)=0$ 
is simple and the entropy of entanglement is given by
\begin{equation}
E(|\psi\rangle)=h\left(\frac{1+\sqrt{1-[C_{2}(|\psi\rangle)]^{2}}}{2}\right)\;,
\end{equation} 
where $h(x)=-x\log x-(1-x)\log (1-x)$. This formula holds for mixed states
where the concurrence for mixed states is defined in Eq.~(\ref{CMM}) and the LHS is 
replaced by the entanglement of formation~\cite{Woo98}.  

For $d=3$, the solutions to the cubic equation $f_{\lambda}(x)=0$ 
are more complicated (although possible) and the entropy of entanglement 
is given by
\begin{align}
&E(|\psi\rangle) =H
\Big({1\over 3}+\frac{2}{3}\sqrt{1-[C_{2}(|\psi\rangle)]^{2}}
\cos\left(\theta / 3\right),\nonumber\\
& \;\;\;\;\;\;\;\;\;\;\;\;\;\;\;\;{1\over 3}+\frac{2}{3}\sqrt{1-[C_{2}(|\psi\rangle)]^{2}}
\cos\big((\theta +2\pi)/ 3\big)\Big)\;;\nonumber\\
&\cos\theta\equiv\frac{1-{3\over 2}[C_{2}(|\psi\rangle)]^{2}
+{1\over 2}[C_{3}(|\psi\rangle)]^{3}}
{\left(1-[C_{2}(|\psi\rangle)]^{2}\right)^{3/2}}\;,
\label{qutrit}
\end{align}
where $H(x,y)=-x\log x-y\log y-(1-x-y)\log (1-x-y)$. Similarly, for $k=4$, it is possible
to find the solutions to the quartic equation $f_{\lambda}(x)=0$ and express the
entropy of entanglement in terms of the concurrence monotones.   

The analytical expression for $C_{k}(|\psi\rangle)$ 
in terms of the reduced density matrix $\rho_{r}\equiv {\rm Tr} _{{}_{B}}|\psi\rangle \langle \psi|$
is given by:
\begin{align}
& C_{k}(|\psi\rangle)=\nonumber\\
& d\left[\frac{1}{{d\choose k}}
\sum_{\{N_{m}\}}(-1)^{k-\sum_{m=1}^{k} N_{m}}\prod_{m=1}^{k}\frac{1}{N_{m}!}
\left(\frac{{\rm Tr}\rho _{r} ^{m}}{m}\right)^{N_{m}}\right]^{1/k}\;,
\label{Powerful}
\end{align}
where the sum is taken over all the non-negative integers $N_{1},N_{2},...,N_{k}$
that satisfy the constraint $N_{1}+2N_{2}+...+kN_{k}=k$. This expression
(see also~\cite{Byr03,Kim03})
follows directly from multinomial formulas given in~\cite{AB64}. 
As an example, for $k=2,3,4$ Eq.~(\ref{Powerful}) gives
\begin{align}
C_{2}(|\psi\rangle) & =\sqrt{{d\over d-1}\left(1-{\rm Tr}\rho_{r}^{2}\right)}
\nonumber\\
C_{3}(|\psi\rangle) & =\left[{d^{2}\over (d-1)(d-2)}\left(1-3{\rm Tr}\rho_{r}^{2}
+2{\rm Tr}\rho _{r}^{3}\right)\right]^{1/3}\nonumber\\ 
C_{4}(|\psi\rangle) & =\Big[{d^{3}\over (d-1)(d-2)(d-3)}\times\nonumber\\
&\left(1-6{\rm Tr}\rho_{r}^{2}
+8{\rm Tr}\rho _{r}^{3}-6{\rm Tr}\rho_{r}^{4}
+3({\rm Tr}\rho_{r}^{2})^{2}\right)\Big]^{1/4}\;.
\label{ex}
\end{align}
We can see that for $k=2$ Eq.~(\ref{Powerful}) is reduced to the 
expression for the concurrence given in~\cite{Run01}. Note also that 
$C_{k}(|\psi\rangle)=0$ if $k$ is greater then the 
Schmidt number of $|\psi\rangle$.

\subsubsection*{The $G$-concurrence monotone}

The last member of the family $C_{k=d}$ is of a particular importance and we denote
it by $G_{d}$ since it is the {\em geometric mean} of the Schmidt numbers
\begin{equation} 
G_{d}(|\psi\rangle)\equiv C_{k=d}(|\psi\rangle)
=d(\lambda_{0}\lambda_{1}\cdots\lambda_{d-1})^{1/d}.
\end{equation}
Note that for $d=2$ the $G$-concurrence coincides with the original definition
of concurrence given by Hill and Wootters~\cite{HW97}.
 
The $G$-concurrence has several interesting features:\\ 
$\bullet$ {\em A computational manageable measure of entanglement}: 
for the $d\times d$ bipartite pure state
$|\psi\rangle$ in Eq.~(\ref{state}), the
$G$-concurrence is given simply by~\footnote{Despite the simple expression in Eq.~(\ref{deta})
for pure states, the convex roof for the G-concurrence on mixed states is yet unknown.
On the other hand, the multi-partite, two level generalizations of concurrence~\cite{Uhl00} 
do admit an explicit formula for the convex roof.}  
({\it cf} Eq.~(\ref{123}))
\begin{equation}
G_{d}(|\psi\rangle) =d\left[{\rm Det}\left(A^{\dag}A\right)\right]^{1/d}\;,
\label{deta}
\end{equation}
where the matrix elements of $A$ are $a_{ij}$.\\
$\bullet$ {\em Multiplicativity}: first, given a 
$d_{1}\times d_{1}$ ($d_{2}\times d_{2}$)
bipartite entangled state, $|\psi _{1}\rangle$ ($|\psi _{2}\rangle$),
we have
\begin{equation} 
G_{d_{1}d_{2}}(|\psi_{1}\rangle\otimes|\psi_{2}\rangle)
=G_{d_{1}}(|\psi_{1}\rangle)G_{d_{2}}(|\psi_{2}\rangle)\;.
\end{equation} 
Note that although in both sides of the equation above we take the geometric 
means of the Schmidt numbers of the relevant states, $G_{d_{1}d_{2}}=C_{k=d_{1}d_{2}}$
is not the {\em same} measure of entanglement as $G_{d_{1}}=C_{k=d_{1}}$~\footnote{
For example, if $d_{1}=d_{2}=2$ then it is clear from 
Eq.~(\ref{ex}) that $C_{k=d_{1}=2}$ is a completely different measure then 
$C_{k=d_{1}d_{2}=4}$}. Second, given a bipartite state 
$|\psi\rangle\in {\cal H}_{A}\otimes{\cal H}_{B}$, a complex number $c$ and 
operators (complex matrices) $\hat{A}\in{\cal H}_{A}$ and $\hat{B}\in {\cal H}_{B}$
we have~\footnote{The determinant of an operator, like its trace, is basis independent.}
\begin{align}
G_{d}(c|\psi\rangle) & = |c|^{2}G_{d}(|\psi\rangle)\label{m1}\\
G_{d}\left(\hat{A}\otimes\hat{B}|\psi\rangle\right) & =
\left|{\rm Det}\left(\hat{A}\right)\right|^{2/d}\left|{\rm Det}\left(\hat{B}\right)
\right|^{2/d}G_{d}(|\psi\rangle)\;,\label{m2}
\end{align}   
where we have used Eq.~(\ref{deta}).\\
$\bullet$ {\em A lower bound}: the $G$-concurrence monotone provides a lower bound for all the other
concurrence monotones. First, for {\em pure} bipartite states we have the inequalities 
({\it cf} p.224 in~\cite{Majo})
\begin{equation}
[C_{2}(|\psi\rangle)]^{2}\geq[C_{3}(|\psi\rangle)]^{3}\geq\cdots\geq
[C_{d}(|\psi\rangle)]^{d}\equiv[G_{d}(|\psi\rangle)]^{d}\;.\label{gb1}
\end{equation}
Second, given a {\em mixed} bipartite state $\rho$ we have~\footnote{For pure states, 
Eq.~(\ref{gb2}) follows from the geometric-arithmetic inequality, and for mixed states 
from the convex roof extension.}    
\begin{equation}
G_{d}(\rho)\leq C_{k}(\rho)\;\;\;\forall\;\;k=1,2,...,d.
\label{gb2}
\end{equation} 
Note that the relations in Eqs.~(\ref{gb1},\ref{gb2}) may be useful in finding 
lower bounds on measures of entanglement such
as entanglement of formation.
In addition, as we will see in the following section, the $G$-concurrence monotone
plays a central role in tripartite RED protocols.

\section{Remote Entanglement Distribution }

As mentioned in the introduction, shared bipartite entanglement is a crucial shared resource for 
many quantum information tasks
such as teleportation~\cite{Ben93}, entanglement
swapping~\cite{Zuk93}, 
and remote state preparation (RSP)
\cite{Ben01,Shi02,Leu03,Ye04} that are employed
in quantum information protocols. 

Remote preparation of bipartite entangled states~\cite{GS04} (RPBES)
is another important quantum information task in which a quantum network (QNet) have
a single supplier (named ``Sapna'') who shares entangled states with nodes via quantum channels,
then performs LOCC to produce pairwise entangled states between any two nodes, say,
Alice and Bob. A crucial feature of RPBES is that Alice and Bob end up sharing
a {\em unique} bipartite entangled state. A more general scheme, in which Alice and Bob
end up sharing a {\em distribution} of entangled states is called remote entanglement
distribution~\cite{GS04}(RED). 

The scheme for tripartite RED, introduced in~\cite{GS04}, commences with a four-way 
shared state, $\hat{\rho} _{1234}=\hat{\rho}_{12}\otimes\hat{\rho} _{34}$
with $\hat{\rho} _{12}$ and $\hat{\rho} _{34}$ bipartite entangled states,
and with Sapna (the supplier) holding shares 2 and 3, and Alice and Bob holding shares 1
and 4, respectively. Each share has a corresponding $d$-dimensional Hilbert space.
The three parties Alice, Bob and Sapna perform LOCC 
to create a set of outcomes 
\begin{equation}
\label{eq:outcome}
\mathcal{O}\equiv\{\hat{\sigma}_{14}^j=\text{Tr}_{23}\hat\sigma^j_{1234},
        Q_j;j=1,\ldots,s\}
\end{equation}
with $Q_j$ the probability that Alice and Bob
share the mixed state $\hat{\sigma}_{14}^j$ which is
obtained by reducing the four-way shared state $\hat\sigma^j_{1234}$ over Sapna's shares. 
In general RED, the states $\{\hat{\sigma}_{14}^j\}$
may be inequivalent under LOCC
whereas in RBESP the states $\hat{\sigma}_{14}^j$
shared by Alice and Bob must be equivalent under LOCC, so Alice and Bob can
always transform $\hat{\sigma}_{14}^j$ into a unique entangled state (i.e. independent on $j$)
via LOCC.

In this section, we address the issue of which distributions of states, $\mathcal{O}$, 
can or cannot be created via LOCC by Alice, Bob and Sapna. The $G$-concurrence 
monotone plays a major role in the following theorem that establishes which 
distributions of states cannot be produced by RED.

\begin{theorem}
If Alice, Bob and Sapna perform LOCC on the initial 4-qudit state 
$\hat{\rho}_{12}\otimes\hat{\rho}_{34}$
with $O$ (in Eq.~(\ref{eq:outcome})) 
the resultant distribution of states shared between Alice and Bob, then
\begin{equation}
G_{14}\equiv\sum_{j=1}^{s}Q_{j}G_{d}(\hat{\sigma}_{14}^{j})\;\leq\;G_{12}G_{34},
\label{theorem}
\end{equation}
with $G_{12}\equiv G_{d}(\hat{\rho} _{12})$ and
$G_{34}\equiv G_{d}(\hat{\rho} _{34})$.
\end{theorem}
(In the next subsection we will show that the equality in the above
equation can always be achieved by RBESP if $\hat{\rho}_{12}$ and $\hat{\rho}_{34}$
are pure.) 

\textbf{Proof:} Let us write $\hat{\rho} _{12}$ and $\hat{\rho} _{34}$ in their
\emph{optimal} decompositions
\begin{equation}
\hat{\rho}_{12} = \sum_{l=0}^{d^{2}-1} p_{l}|\psi ^{(l)}\rangle_{12}\langle\psi ^{(l)}|
\;,\;\;\hat{\rho}_{34} = \sum_{l=0}^{d^{2}-1} q_{l}|\chi ^{(l)}\rangle
_{34}\langle\chi ^{(l)}|;
\label{initial}
\end{equation}
we can always choose optimal
decompositions with no more then $d^2$ elements~\footnote{Although the optimal decompositions 
in Eq.~(\ref{initial}) are taken with $d^{2}$ elements, it is not necessary for the proof; 
we could instead write the optimal decompositions with any number of elements.}. 
The states $|\psi ^{(l)}\rangle _{12}$ and 
$|\chi ^{(l)}\rangle _{34}$ are given in their 
Schmidt decomposition:
\begin{align}
|\psi ^{(l)}\rangle _{12} & = \sum _{k=0}^{d-1}\sqrt{\lambda _{k}^{(l)}}|k^{(l)}
k^{(l)}\rangle _{12}\nonumber\\
|\chi ^{(l)}\rangle _{34} & = \sum _{k=0}^{d-1}\sqrt{\eta _{k}^{(l)}}|k^{(l)}
k^{(l)}\rangle _{34}\;,
\end{align}
with $\lambda _{k}^{(l)}$ and $\eta _{k}^{(l)}$ 
the Schmidt coefficients of $|\psi ^{(l)}\rangle _{12}$ and $|\chi ^{(l)}\rangle
_{34}$, respectively. The index $l$ in the states $\{|k^{(l)}\rangle
_{i}\}$ represents $d^2$ different bases for each
system $i=1,2,3,4$. Note that in this notation 
\begin{align}
G_{12} & =d\sum _{l=0}^{d^{2}-1}p_{l}\left(\lambda _{0}^{(l)}\lambda _{1}^{(l)}
\cdots\lambda _{d-1}^{(l)}\right)^{1 \over d}\nonumber\\
G_{34} & =d\sum _{l=0}^{d^{2}-1}q_{l}\left(\eta _{0}^{(l)}\eta _{1}^{(l)}
\cdots\eta _{d-1}^{(l)}\right)^{1 \over d}\;.
\label{G12}  
\end{align}

Since the entanglement between Alice and Bob remains zero
unless Sapna perform a measurement, we assume that the first 
measurement is performed by Sapna and is described
by the Kraus operators $\hat{M}^{(j)}$ and their components
\begin{equation}
M^{(j,ll')}_{mm',kk'}
\equiv {}_{23}\langle m^{(l)} m'^{(l')}|\hat{M}^{(j)}|k^{(l)}k'^{(l')}\rangle_{23}\;,
\end{equation}
with $k,k',m,m'=0,1$ and $l,l'=1,2,3,4$.

The probability to obtain an outcome $j$ is thus 
\begin{align}
Q_{j} & \equiv \text{Tr}( \hat{M}^{(j)}\hat{\rho}_{12}\otimes
\hat{\rho}_{23} \hat{M}^{(j)\dag})
& =\sum_{l=0}^{d^{2}-1}\sum_{l'=0}^{d^{2}-1}p_{l}q_{l'}N^{(j,ll')}, 
\end{align}
with
$N^{(j,ll')}\equiv\sum_{m,m'}r^{(j,ll')}_{mm'}$ and 
\begin{equation}
r^{(j,ll')}_{mm'} \equiv  \sum _{k,k'}\lambda _{k}^{(l)}\eta_{k'}^{(l')}
|M^{(j,ll')}_{kk',mm'}|^{2}\;.
\end{equation}
The density matrix shared between Alice,
Bob and Sapna after outcome $j$ occurs is
\begin{equation}
\hat{\sigma}_{1234}^{j}=\frac{1}{Q_{j}}\sum_{l,l'}p_{l}q_{l'}N^{(j,ll')}|\Phi^{(j,ll')}\rangle
_{1234}\langle \Phi^{(j,ll')}|\;,
\end{equation}
where
\begin{align}
|\Phi^{(j,ll')}&\rangle _{1234}=\frac{1}{\sqrt{N^{(j,ll')}}}
\sum _{k,k'}\sum_{m,m'}\sqrt{\lambda_{k}^{(l)}\eta_{k'}^{(l')}}\nonumber\\
&\times M^{(j,ll')}_{kk',mm'}
|k^{(l)}k'^{(l')}\rangle_{14}|m^{(l)} m'^{(l')}\rangle _{23}
\end{align}
Tracing over Sapna's subsystems yields
\begin{equation}
\hat{\sigma}_{14}^j=\frac{1}{Q_j}\sum_{l,l'}\sum_{m,m'}p_lq_{l'}r^{(j,ll')}_{mm'}
        |\phi^{(j,ll')}_{mm'}\rangle _{14}\langle \phi^{(j,ll')}_{mm'}|\;, 
\label{deco}
\end{equation}  
where
\begin{equation}
|\phi^{(j,ll')}_{mm'}\rangle _{14} \equiv 
\frac{1}{\sqrt{r^{(j,ll')}_{mm'}}} 
\sum _{k,k'}\sqrt{\lambda _{k}^{(l)}\eta
  _{k'}^{(l')}}M^{(j,ll')}_{kk',mm'}|k^{(l)}
k'^{(l')}\rangle_{14}\;.\label{phi}
\end{equation}
From the definition of the G-concurrence for mixed states (i.e. the convex roof extension),
it follows that
$G_{d}(\hat{\sigma}_{14}^{j})$ cannot exceed
the average of the $G$-concurrence over the decomposition in Eq.~(\ref{deco}). 
Thus,
\begin{equation}
G_{d}\left(\hat{\sigma}_{14}^j\right)\leq \frac{1}{Q_j}\sum_{l,l'}
\sum_{m,m'}p_lq_{l'}r^{(j,ll')}_{mm'}
G\left(|\phi^{(j,ll')}_{mm'}\rangle _{14}\right).
\label{bnd}
\end{equation}
Using Eq.~(\ref{deta}) we find
\begin{equation}
G\left(|\phi^{(j,ll')}_{mm'}\rangle _{14}\right) =
\frac{d\left(\prod_{k=0}^{d-1}\lambda _{k}^{(l)}\eta _{k}^{(l')}\right)^{1/d}
\left|{\rm Det}\left({\cal M}^{(j,ll')}_{mm'}\right)\right|^{2/d}}{r^{(j,ll')}_{mm'}}\;,
\end{equation}
where the $d^{2}$ elements of each matrix ${\cal M}^{(j,ll')}_{mm'}$
are $M^{(j,ll')}_{kk',mm'}$.
Thus, substituting this result in Eq.~(\ref{bnd}) yields
\begin{align}
&  G_{14}\equiv \sum_{j=1}^{s}Q_{j}G\left(\hat{\sigma}_{14}^{j}\right)
\leq d\sum _{l,l'}p_{l}q_{l'}\left(\prod_{k=0}^{d-1}\lambda _{k}^{(l)}\eta _{k}^{(l')}\right)^{1/d}
\nonumber\\ 
& \times  \sum_{j}\sum_{m,m'}
\left|{\rm Det}\left({\cal M}^{(j,ll')}_{mm'}\right)\right|^{2/d} \;.\label{gg}
\end{align}
Now, from the geometric-arithmetic inequality we have
\begin{align}
\sum_{m,m'}\left|{\rm Det}\left({\cal M}^{(j,ll')}_{mm'}\right)\right|^{2/d}
& \leq\frac{1}{d}\sum_{m,m'}{\rm Tr}\left({\cal M}^{(j,ll')\dag}_{mm'}
{\cal M}^{(j,ll')}_{mm'}\right)\nonumber\\
&=\frac{1}{d}{\rm Tr}(\hat{M}^{(j)\dag}\hat{M}^{(j)}).
\end{align}
Hence, from Eq.~(\ref{gg}) and Eq.~(\ref{G12}) we get 
\begin{equation}
G_{14}\leq\frac{1}{d^{2}} 
G_{12}G_{34}\sum_{j}{\rm Tr}(\hat{M}^{(j)\dag}\hat{M}^{(j)})\;.
\label{box}
\end{equation}
Thus, from the completeness relation,
$\sum_{j}\hat{M}^{(j)\dag}\hat{M}^{(j)}=I$, we obtain Eq.~(\ref{theorem}). 

Consider now the following LOCC: after Sapna's first measurement, she  
sends the result $j$ to Alice and Bob. Based on this result, Alice then performs a 
measurement represented by the Kraus operators $\hat{A}^{(k)}_{j}$ and sends
the result $k$ to Bob and Sapna. Based on the results $j,k$ from Sapna and Alice,
Bob performs a measurement represented by the Kraus operators $\hat{B}^{(n)}_{jk}$
and send the result $n$ to Sapna. In the last step of this scheme, Sapna performs
a second measurement with Kraus operators denoted by $\hat{F}_{jkn}^{(j)}$ and send 
the result $i$ to Alice and Bob. The final distribution of entangled
states shared between Alice and Bob is denoted by $\{N_{jkni},\sigma
_{14}^{jkni}\}$, where $N_{jkni}$ is the probability for outcome
$j,k,n,i$ and $\hat{\sigma}_{14}^{jkni}={\rm Tr}_{{}_{23}}\hat{\sigma} ^{jkni}_{1234}$ 
with 
\begin{align}
\hat{\sigma} ^{jkni}_{1234} =\frac{1}{N_{jkni}}&\left(\hat{A}^{(k)}_{j} \otimes
\hat{F}^{(i)}_{jkn}\hat{M}^{(j)} \otimes\hat{B}^{(n)}_{jk}\right)
\left[\hat{\rho}_{12}
\otimes\hat{\rho}_{34}\right]\nonumber\\
& \;\;\;\;\;\;\left(\hat{A}^{(k)}_{j}\otimes
\hat{F}^{(i)}_{jkn}\hat{M}^{(j)}\otimes\hat{B}^{(n)}_{jk}\right)^{\dag}\;.
\end{align}

Since the G-concurrence of any bipartite state
satisfies Eq.~(\ref{m2}), the analog of Eq.~(\ref{box}) for this LOCC protocol
is therefore,
\begin{align}
&  G_{14}\equiv\nonumber\\
& \sum_{j,k,n,i}N_{jkni}G\left(\hat{\sigma}_{14}^{jkni}\right)
\leq\frac{1}{d^{2}}G_{12}G_{34}\sum_{j,k}\left|{\rm Det}(\hat{A}_{j}^{(k)})\right|^{2/d}\nonumber\\
& \times \sum_{n}
\left|{\rm Det}(\hat{B}_{jk}^{(n)})\right|^{2/d}\sum_{i}{\rm Tr}
\left(\hat{M}^{(j)\dag}\hat{F}^{(i)\dag}_{jkn}
\hat{F}^{(i)}_{jkn}\hat{M}^{(j)}\right)\;.
\label{nonbasic}
\end{align}

Moreover, from the geometric-arithmetic inequality we have
\begin{equation}
\sum_{n}\left|{\rm Det}(\hat{B}_{jk}^{(n)})\right|^{2/d}\leq 
\frac{1}{d}\sum _{n}{\rm
  Tr}\hat{B}_{jk}^{(n)\dag}\hat{B}_{jk}^{(n)}=1
\end{equation}
and a similar
relation for $\hat{A}_{j}^{(k)}$.
These results, together with
the completeness relation
$\sum_{i}\hat{F}^{(i)\dag}_{jkn}\hat{F}^{(i)}_{jkn}=1$,
lead us back to Eq.~(\ref{box}).
As we can see, all operations that are performed by Alice, Bob and Sapna
after the first measurement by Sapna cannot increase the bound on $C_{14}\;\Box$.  

Theorem~1 concerns one supplier and two nodes, but in fact applies to one supplier
and \emph{any} pair of nodes; thus, the result of Theorem~1 is applicable to an arbitrarily
large QNet with one supplier and many nodes. In fact Theorem~1 can be extended to
more than one supplier, as stated in the following corollary.

{\bf Corollary}: Consider an align chain of $N$ mixed bipartite states, 
$\rho _{0,1},\;\rho _{1,2},...,\rho_{N-1,N}$, where the state $\rho_{k-1,k}$
($k=1,2,...,N$) is shared between party $k-1$ and party $k$. If  
the $N+1$ parties perform LOCC on the initial state 
$\rho _{0,1}\otimes\rho _{1,2}\otimes\cdots\otimes\rho_{N-1,N}$ with 
the resultant distribution of states between party $0$ and $N$ denoted by
$\{P_{j},\;\hat{\sigma}_{0N}^{j}\}$ ($P_{j}$ is the probability to have 
the state $\hat{\sigma}_{0N}^{j}$), then
\begin{equation}
G_{0N}\equiv\sum_{j}P_{j}G_{d}(\hat{\sigma}_{0N}^{j})\;\leq\;G_{01}G_{12}\cdots G_{N-1_{\;}N}\;,
\label{cor}
\end{equation}    
with $G_{k-1_{\;}k}\equiv G_{d}(\rho_{k-1,k})$ ($k=1,2,...,N$).

Theorem 1 and its corollary suggest an operational interpretation of the G-concurrence
as a form of \emph{entanglement capacity}.  
In the following subsection we show that if both $\hat{\rho}_{12}$ and $\hat{\rho}_{23}$
are $d\times d$-dimensional {\em pure} states, than the equality in 
Eqs.~(\ref{theorem},\ref{cor}) can always
be achieved.  
 
\subsection*{An optimal protocol for RPBES}

In this section we show that by LOCC Sapna can prepare a bipartite pure state
between Alice and Bob with {\em any} value of the concurrence monotone $G$
which is less or equal to $G_{12}G_{34}$. For this purpose,
we introduce the protocol for RBESP that has been
first introduced in~\cite{GS04}. In this protocol the supplier Sapna
shares the initial $(d\times d)$-dimensional pure states
$|\psi\rangle_{12}  =\sum_{k=0}^{d-1}\sqrt{\lambda _{k}}|kk\rangle _{12}$ and
$|\chi\rangle_{34}  =\sum_{k=0}^{d-1}\sqrt{\eta _{k}}|kk\rangle _{34}$
(which are expressed in the Schmidt decomposition) with Alice and Bob, 
respectively.

The steps of the protocol are as follows:\\
\textbf{(i)}~Sapna performs a projective measurement
\begin{equation}
\hat{P}^{(j,j')}=|P^{(j,j')}\rangle_{23}\langle P^{(j,j')}|,\,j,j'=0,1,...,d-1,
\label{eq:proj}
\end{equation}
with
\begin{equation}
|P^{(j,j')}\rangle_{23} \equiv   
 \frac{1}{d}\sum_{m,m'=0}^{d-1}\text{e}^{i\left[\frac{2\pi}{d^{2}}(dj+j')(dm+m')+\theta
 _{mm'}\right]}|mm'\rangle _{23}\;, 
\label{meas}
\end{equation}
with $\theta _{mm'}\in\mathbb{R}$ chosen freely. Note that the $d^2$
states $|P^{(j,j')}\rangle _{23}$ are orthonormal, regardless of the
choice of $\theta _{mm'}$.\\
\textbf{(ii)}~After the outcomes $j,j'$ have been obtained, the state of the system
can be written as $|P^{(j,j')}\rangle _{23}|\phi ^{(j,j')}\rangle _{14}$, where
\begin{align}
|\phi ^{(j,j')}\rangle _{14} & =  \sum _{m=0}^{d-1}\sum_{m'=0}^{d-1}
\sqrt{\lambda _{m}\eta _{m'}}\nonumber\\
& \times  \text{e}^{-i\left[\frac{2\pi}{d^{2}}(dj+j')(dm+m')+\theta _{mm'}\right]}|mm'\rangle _{14}\;.
\end{align}
\textbf{(iii)}~Sapna sends the results $j$ and $j'$ to Bob ($2\log_{2}d$ bits of
information) and 
the result $j'$ ($\log_{2}d$ bits of information) to Alice. 
Bob then performs the unitary operation
\begin{equation}
\hat{U}_{b}^{(j,j')}|m'\rangle
_{4}=\exp\left(i\frac{2\pi}{d^{2}}(dj+j')m'\right)|m'\rangle_4\;,
\label{Ch}
\end{equation}
and Alice performs the unitary operation
\begin{equation}
\hat{U}_{a}^{(j')}|m\rangle_1=\exp\left(i\frac{2\pi}{d}j'm\right)|m\rangle _1\;.
\label{Al}
\end{equation}
\textbf{(iv)}~The final state shared between Alice and Bob is
\begin{equation}
|F\rangle _{14}=\sum_{m=0}^{d-1}\sum_{m'=0}^{d-1}\exp\left
(-i\theta _{mm'}\right)\sqrt{\lambda_{m}\eta _{m'}}|mm'\rangle _{14},
\label{final}
\end{equation}
(which is separable for $\theta _{mm'}=0$). 

We will show now, that by choosing the phases $\theta _{mm'}$ appropriately, 
Sapna can prepare the state $|F\rangle _{14}$ with {\em any} value of 
$G(|F\rangle _{14})$ in the range $[0,G_{12}G_{34}]$. For this purpose,
we define the square ($d\times d$) complex matrix $A$ with elements
$a_{mm'}=\sqrt{\lambda_{m}\eta _{m'}}\exp\left
(-i\theta _{mm'}\right)$. Thus,
\begin{align}
G(|F\rangle _{14}) & =d\left[{\rm Det}\left(A^{\dag}A\right)\right]^{1/d}
\nonumber\\
& =G_{12}G_{34}\left[{\rm Det}\left(V^{\dag}V\right)\right]^{1/d}\;,
\end{align} 
where $G_{12}=d(\lambda_{0}\lambda_{1}\cdots\lambda_{d-1})^{1/d}$
$G_{34}=d(\eta_{0}\eta_{1}\cdots\eta_{d-1})^{1/d}$ and the matrix elements
of $V$ are $v_{mm'}=\exp\left (-i\theta _{mm'}\right)/\sqrt{d}$. Note that
for the choice $\theta _{mm'}=2\pi mm'/d$ the matrix $V$ is unitary and therefore
$G(|F\rangle _{14})=G_{12}G_{34}$. For other choices of $\theta _{mm'}$, Sapna can
prepare the final state $|F\rangle _{14}$ with any value of the $G$-concurrence
monotone in the range $[0,G_{12}G_{34}]$. 

It is important to emphasize here that
the choice $\theta _{mm'}=2\pi mm'/d$ maximizes {\em only} the $G$-concurrence.
In fact, for other measures of entanglement the values of $\theta _{mm'}$ that
maximize the entanglement depend explicitly on the Schmidt numbers $\lambda_{m}$
and $\eta _{m}$.
For example, the concurrence monotone $C_{k=2}$ of the final state $|F\rangle _{14}$
is
\begin{align}
C_{2}\left(|F\rangle _{14}\right) & =  
2{\Big\{}\sum_{k>k'}\sum_{m>m'}\lambda _{k}\lambda _{k'}\eta
  _{m}\eta _{m'}\nonumber\\
& \times 
\left|\text{e}^{i(\theta _{km}+\theta _{k'm'})}
-\text{e}^{i(\theta _{km'}+\theta _{k'm})}\right|^{2}{\Big\}}^{1/2}.
\label{eq:CF}
\end{align}
Thus, in this case we see that the values of $\theta _{km}$ that maximize 
$C_{2}\left(|F\rangle _{14}\right)$ depend explicitly on the Schmidt 
coefficients $\lambda_{k}$ and $\eta _{m}$.

\section{Summary and conclusions} 

In summary, we have introduced a family of entanglement monotones
that extend the definition of concurrence. We have shown that for 
a finite number of copies of pure states (i.e. the deterministic case)
the family characterizes completely the non-local resource. We have also discussed
the advantage of the concurrence monotones over other measures of entanglement
(such as the entropy of entanglement, the Renyi entropies, etc.) and showed that 
for a given bipartite state, $|\psi\rangle =\sum_{ij}a_{ij}|i\rangle |j\rangle$,
the concurrence monotones can always be expressed analytically in terms of the 
coefficients $a_{ij}$. We also gave an analytical expression of the concurrence 
monotones (for pure states) in terms of the reduced density matrix 
(see Eq.~(\ref{Powerful})).  

We then discussed a particular member of the family which we called the $G$-concurrence.
The $G$-concurrence for pure states is defined as the geometric mean of the Schmidt numbers.
It has several unique properties that makes it extremely useful. In particular, 
we have proved a powerful theorem that establishes an upper bound on the amount of
$G$-concurrence that can be created between two single-qudit nodes of quantum networks
by means of RED. The theorem also suggests an operational interpretation of the G-concurrence
as a type of entanglement capacity. We have proved that it is always possible to saturate the 
$G$-concurrence bound in the theorem if both of the entangled states are pure, and
also suggested an operational interpretation of the G-concurrence
as a type of entanglement capacity. An 
open question is left if it is possible to saturate the bound when the states are mixed.

The concurrence monotones are defined in terms of the symmetric functions of the Schmidt 
numbers (see Eq.~(\ref{sym})). These symmetric functions have many interesting mathematical
properties which were not introduced here 
(some of the properties can be found in~\cite{Majo}) and which are related to the field
of majorization. Thus, we believe that further investigations of these monotones will
contribute to our understanding of entanglement.

\textbf{Acknowledgments:} I would like to extend my sincere gratitude
to Barry Sanders, for fruitful discussions and for reviewing this work 
in its preliminary stages. The author acknowledges support by
the Killam Trust, the DARPA QuIST program under contract F49620-02-C-0010, 
and the National Science Foundation (NSF) under grant ECS-0202087.

\end{document}